\documentclass[conference]{IEEEtran}
\usepackage{diagbox} 
\usepackage{graphicx}
\usepackage{epstopdf}
\usepackage{psfrag}
\usepackage{subfigure}
\usepackage{url}
\usepackage{stfloats}
\usepackage{amsfonts,amssymb,amsmath,bm,paralist,cite,ifthen,color}
\usepackage{array}
\usepackage{caption}
\usepackage{calc}
\usepackage{enumerate}
\usepackage[ruled]{algorithm2e}
\usepackage{array}
\usepackage{float}
\usepackage[colorlinks,
            linkcolor=black,
            anchorcolor=black,
            citecolor=black]{hyperref}

\newtheorem{remark}{Remark}

\graphicspath{{Figures/}}

\newcommand\nthis{\addtocounter{equation}{1}\tag{\theequation}}
\newcommand\Cc{\ensuremath{\mathcal{C}}}
\newcommand\Nc{\ensuremath{\mathcal{N}}}
\newcommand\Oc{\ensuremath{\mathcal{O}}}
\newcommand\Ac{\ensuremath{\mathcal{A}}}
\newcommand\Bc{\ensuremath{\mathcal{B}}}
\newcommand\Uc{\ensuremath{\mathcal{U}}}
\newcommand\Lc{\ensuremath{\mathcal{L}}}
\newcommand\Wc{\ensuremath{\mathcal{W}}}

\newcommand\xb{\ensuremath{{\bm x}}}
\newcommand\ab{\ensuremath{{\bm a}}}
\newcommand\eb{\ensuremath{{\bm e}}}
\newcommand\nb{\ensuremath{{\bm n}}}
\newcommand\yb{\ensuremath{{\bm y}}}
\newcommand\ssb{\ensuremath{{\bm s}}}
\newcommand\ub{\ensuremath{{\bm u}}}

\newcommand\wb{\ensuremath{{\bm w}}}

\newcommand\Bb{\ensuremath{{\bm B}}}

\newcommand\Fb{\ensuremath{{\bm F}}}
\newcommand\Ib{\ensuremath{{\bm I}}}

\newcommand\Ub{\ensuremath{{\bm U}}}
\newcommand\Hb{\ensuremath{{\bm H}}}

\newcommand\Vb{\ensuremath{{\bm V}}}
\newcommand\Wb{\ensuremath{{\bm W}}}
\newcommand\Jb{\ensuremath{{\bm J}}}

\newcommand\Gammab{\ensuremath{{\bm \Gamma}}}

\newcommand\SigmaB{\ensuremath{{\bm \Sigma}}}

\newcommand\diag{\ensuremath{{\rm diag}}}

\newcommand\rank{\ensuremath{{\rm rank}}}
\newcommand\vectorize{\ensuremath{{\rm vec}}}

\newcommand\Cs{\ensuremath{{\mathbb{C}}}}
\newcommand\Es{\ensuremath{{\mathbb{E}}}}
\newcommand\Rs{\ensuremath{{\mathbb{R}}}}

\IEEEoverridecommandlockouts

\begin{document}

\title{Switch-based Hybrid Beamforming for Wideband Multi-carrier Communications
\thanks{This work has been supported in part by Academy of Finland under 6Genesis Flagship (grant 318927) and EERA Project (grant 332362).}
}

\author{\IEEEauthorblockN{Mengyuan Ma, Nhan Thanh Nguyen and Markku Juntti}
\IEEEauthorblockA{\textit{Centre for Wireless Communications (CWC), Uninvesity of Oulu, P.O.Box 4500, FI-90014, Finland}\\
Email: \{mengyuan.ma, nhan.nguyen, markku.juntti\}@oulu.fi}
}

\maketitle

\begin{abstract}
Switch-based hybrid beamforming (SW-HBF) architectures are promising for realizing massive multiple-input multiple-output (MIMO) communications systems because of their low cost and low power consumption. In this paper, we study the performance of SW-HBF in a wideband multi-carrier MIMO communication system considering the beam squint effect. We aim at designing the switch-based combiner that maximizes the system spectral efficiency (SE). However, the design problem is challenging because the analog combing matrix elements are binary variables. To overcome this, we propose tabu search-based (TS) SW-HBF schemes that can attain near-optimal performance with reasonable computational complexity. Furthermore, we compare the total power consumption and energy efficiency (EE) of the SW-HBF architecture to those of the phase-shifter-based hybrid beamforming (PS-HBF) architecture. Numerical simulations show that the proposed algorithms can efficiently find near-optimal solutions. Moreover, the SW-HBF scheme can significantly mitigate the beam squint effect and is less affected by the number of subcarriers than PS-HBF. It also provides improved SE and EE performance compared to PS-HBF schemes.
\end{abstract}

\section{Introduction}
Wideband communications systems, with their large utilizable spectrum, are promising to meet the ever-lasting escalating demand for ultra-high-speed data rates of future 6G wireless networks \cite{niu2015survey,jiang2021road}. However, the large numbers of antennas in millimeter wave (mmWave) and Terahertz (THz) communications systems require large numbers of excessively high power-hungry radio frequency (RF) chains. As a result, there could be prohibitive power consumption and cost. Therefore, hybrid beamforming (HBF) is envisioned as a critical technique to realize mmWave and THz communications. It can considerably reduce the number of costly radio frequency chains and maintain the spatial multiplexing gain \cite{gao2021wideband}. 

\begin{figure}[htbp]
   \vspace{-0.3cm}
        \centering
        \subfigure[Phase shifter-based hybrid beamforming.]{\label{fig: PS HBF} \includegraphics[width=0.4\textwidth]{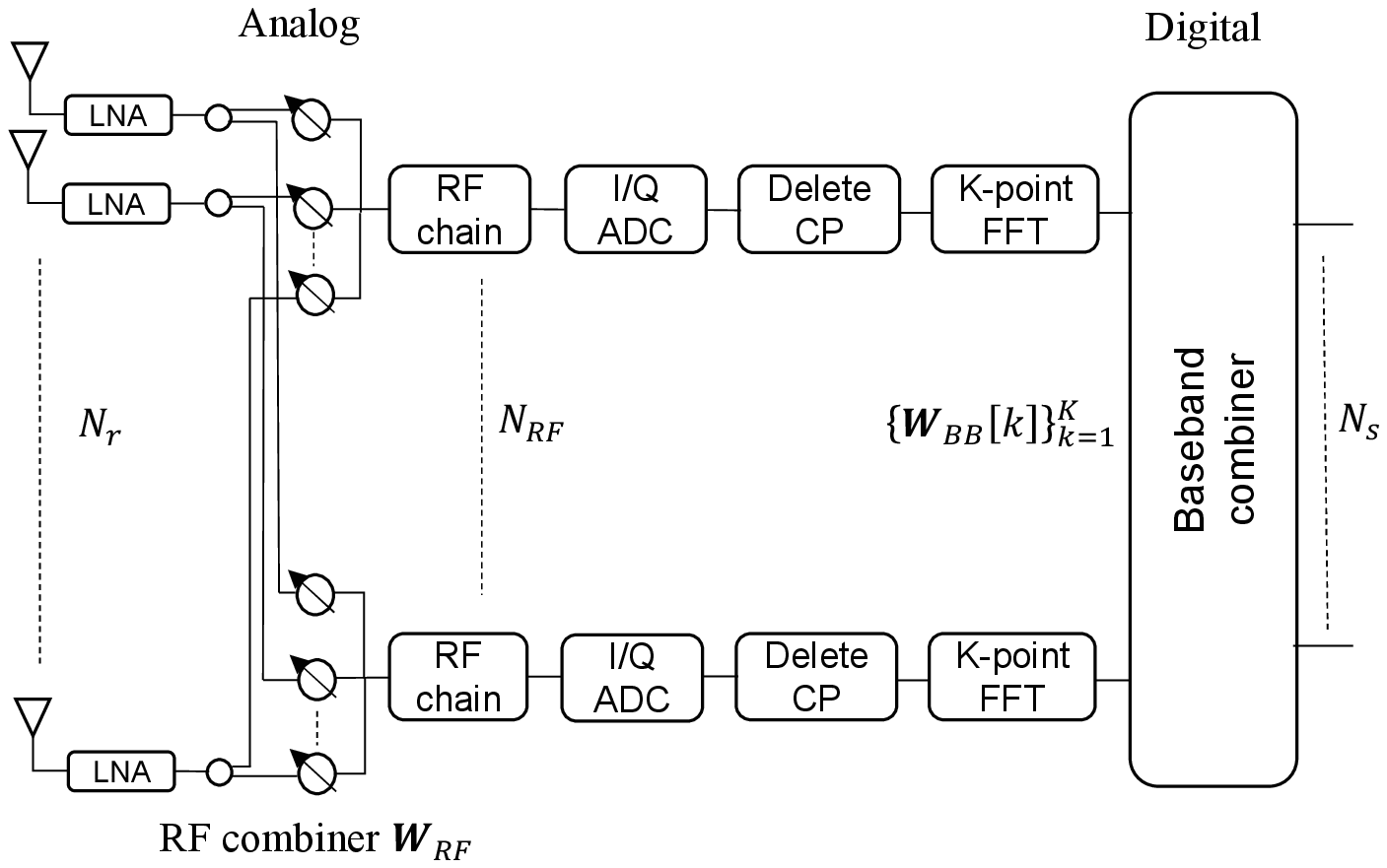}}
        \subfigure[Switch-based hybrid beamforming]{\label{fig: SW HBF} \includegraphics[width=0.4\textwidth]{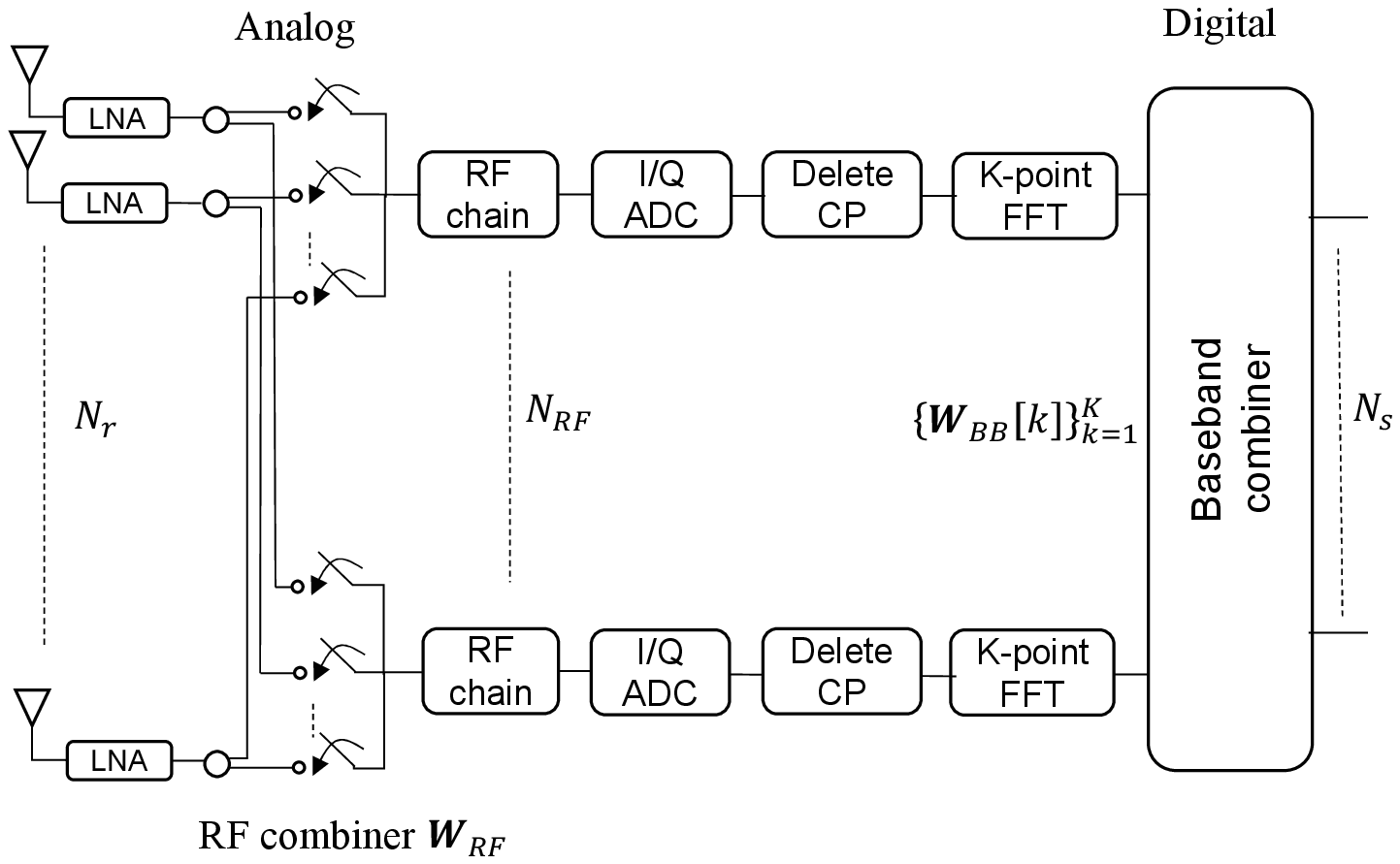}}
        \vspace{-0.3cm}
        \captionsetup{font={small}}
        	\caption{Illustration of PS-based and SW-based hybrid combining structures.}
        	\label{fig:Illustration of hybrid beamforming structure}
        \vspace{-0.3cm}	
  \end{figure}

In HBF, the analog beamformer can be implemented by either soft antenna selection with variable phase-shifters or hard antenna selection  using switch networks \cite{payami2018hybrid} (see Fig.\ \ref{fig:Illustration of hybrid beamforming structure}). Nevertheless, the practical realization of the phase-shifters for high frequencies is not a trivial task \cite{nosrati2019switch}. Moreover, the large number of phase-shifters may require high power consumption, degrading the system energy efficiency (EE). Furthermore, the beam squint effect cannot be neglected in systems employing large bandwidth and large-sized antenna arrays, especially in phase-shift-based HBF (PS-HBF) transceivers \cite{cai2016effect,dai2021delay}. It can significantly degrade the spectral efficiency (SE) in wideband multi-carrier systems. To mitigate the beam squint effect, the true-time-delay (TTD) structure can be embedded into the RF front-end \cite{spoof2020true}, which, however, inevitably causes increased power consumption to the system. In contrast, the frequency-independent switch-based HBF (SW-HBF) is capable of alleviating the beam squint effect without any increase in power consumption. Compared to phase-shifters, switch networks are simple to implement, low-power, and quick to adapt to the channel variations \cite{nosrati2019switch}. Nonetheless, most of the studies on SW-HBF focus on narrowband channel models \cite{mendez2015channel,mendez2016hybrid,jiang2018hybrid,nosrati2019switch}. To the best of the authors' knowledge, SW-HBF in frequency-selective wideband multi-carrier systems has not been thoroughly considered in the literature.
  
In this paper, we investigate the potentials of SW-HBF in overcoming the beam squint effect of a wideband multiple-input multiple-output orthogonal frequency-division multiplexing (MIMO-OFDM) system. Specifically, we aim at designing the SW-based combiner that maximizes the system SE. The design problem is challenging due to the binary variable constraints and the rank constraint of the analog beamformer. To tackle the challenges, we first propose a tabu search (TS) algorithm that is demonstrated to find near-optimal solutions with reasonable complexity. Then, to further refine the solution, we employ a projected gradient ascending (PGA) algorithm to obtain a better initial point for the TS algorithm. Furthermore, we introduce power consumption models for the PS-HBF and the SW-HBF architectures. Finally, intensive simulations are provided to compare the SE and EE of the SW-HBF and PS-HBF schemes. The results show that the former is less affected by the beam squint effect and the number of subcarriers than the latter. Moreover, the proposed TS-based SW-HBF schemes achieve better SE and EE performance than the PS-HBF. 

\section{System Model and Problem Formulation} 
\subsection{Signal Model}
 We consider a point-to-point MIMO-OFDM system where the transmitter is equipped with $N_t$ antennas and perform fully digital beamforming. The receiver employs either PS-HBF or SW-HBF architecture with $N_r$ antennas and $N_{RF}$ RF chains, as illustrated in Figs \ref{fig: PS HBF} and \ref{fig: SW HBF}.  Let $K$ be the number of subcarriers, and let $\ssb_k \in \Cs^{N_s \times 1}$ $(N_s\leq N_{RF})$ be the transmitted symbol vector at the $k$th subcarrier, $\Es\left[\ssb_k\ssb^H_k\right]=\Ib_{N_s},\; k=1,\cdots,K$, where $\Ib_{N_s}$ denotes the $N_s\times N_s$ identity matrix. The transmitted signal vector $\xb_k \in \Cs^{N_t \times 1}$ for each subcarrier is given as
\begin{equation}\label{eq:transmitted signal}
  \xb_k=\Fb_k\ssb_k,
\end{equation}
where $\Fb_k \in \Cs^{N_t \times N_s}$ is the digital precoding matrix. 

At the receiver, the signal vector is first combined by the analog combiner, represented by $\Wb_{RF}\in \Cs ^{N_r \times N_{RF}}$. After discarding the CP and performing $N_{RF}$ $K$-point fast Fourier transforms (FFTs), the combined signal is further processed at frequency domain by low-dimensional baseband combiner $\Wb_{BB}[k] \in \Cs ^{N_{RF} \times N_s}$ for each subcarrier. Finally, the combined signal at the $k$th subcarrier through channel $\Hb_k\in \Cs^{N_r\times N_t}$ is given as
\begin{equation}\label{eq:received signal}
  \yb_k=\Wb^H_k\Hb_k \Fb_k\ssb_k + \Wb^H_k\nb_k,
\end{equation}
where $\Wb_k=\Wb_{RF}\Wb_{BB}[k]$, and $\nb_k\sim \Nc(\boldsymbol{0}, \sigma_n^2\Ib_{N_r})$ is the additive white Gaussian noise vector at the $k$th subcarrier with $\sigma_n^2$ being the noise variance.


\subsection{Beam Squint Effect and Channel Model}\label{sc:channel model}
\subsubsection{Beam Squint Effect}

\begin{figure}[htbp]
   \vspace{-0.3cm}
        \centering	
         \includegraphics[width=0.5\textwidth]{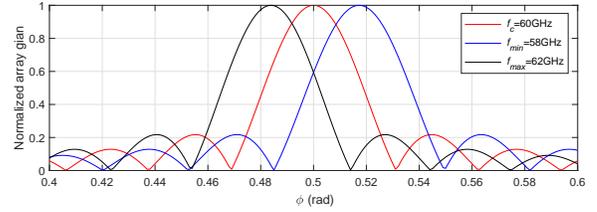}
         \captionsetup{font={small}}
        	\caption{Illustration of beam squint effect in multi-carrier systems with beam focus $\phi_0=\pi/6$, $N=64$ antennas, uniform linear array with antenna sapcing $d_s=\lambda_c/2$, carrier frequency $f_c=60\rm{GHz}$, bandwidth $B=4\rm{GHz}$. }
        	\label{fig: beam squint effect}
        	\vspace{-0.3cm}
  \end{figure}

In the conventional narrowband communications systems with analog beamforming, the phase values of variable phase-shifters are generally optimized for the carrier frequency. This frequency-dependent design incurs the beam squint effect when it is applied to wideband multi-carrier systems \cite{cai2016effect}. Specifically, there may be considerable performance loss for frequencies other than the carrier frequency due to beam patterns of analog beamformers vary with frequencies. Fig.\ \ref{fig: beam squint effect} illustrates the beam patterns as a function of beam focus angle $\phi$ for different frequencies. It can be observed that when the beamformer points to angle $\phi_0=\pi/6$ at carrier frequency $f_c=60\rm{GHz}$, the beamforming gain at other frequencies suffer a significant loss in that their beamforming focus angles squint away from $\pi/6$.
\subsubsection{Wideband Channel Model with Beam Squint Effect}
Since the beam squint effect is essentially induced by the frequency-dependent beam steering vectors, we adopt the modified channel model \cite{li2020dynamic}, which incorporates the frequency dependency into the classical geometric channel model \cite{alkhateeb2016frequency, park2017dynamic, vlachos2019wideband}. Assuming uniform linear array (ULA) is utilized, the $d$-th tap of the channel at frequency $f$ can be modeled as \cite{li2020dynamic}
\begin{equation}\label{eq:d delay channel}
   \Hb_f[d]=\sum_{l=1}^L \alpha_l p\left(d T_s-\tau_l\right) \ab_r\left( \theta_l^r, f\right) \ab_t^H\left( \theta_l^t, f\right), 
\end{equation}
where $L$ is the number of distinct scattering clusters, $\alpha_l \sim \Cc\Nc(0,1)$ and $\tau_l$ are the complex gain and the delay of the $l$th cluster, respectively, $\theta_l^r$ and $\theta_l^t$ represent the angle of arrival (AoA) and angle of departure (AoD) of the $l$th cluster, respectively, and finally $p(\tau)$ denotes the pulse-shaping filter for $T_s$-spaced signaling evaluated at $\tau$ seconds \cite{alkhateeb2016frequency}. The transmit/receive steering vector at frequency $f$ is given by
\begin{equation}\label{eq:array steer vector for MIMO}
   \ab(\theta,f)=\left[1,e^{-j 2 \pi \psi(\theta) \frac{f}{f_c} },\cdots, e^{-j 2 \pi (N-1)  \psi(\theta) \frac{f}{f_c} }  \right]^T, 
\end{equation}
where $N \in \{N_t, N_r \}$ is the number of antennas, $\psi(x) \triangleq \frac{d_s\sin(x)}{\lambda_c}$ with $\lambda_c$ being the wavelength of carrier frequency, and $d_s$ being the antenna spacing distance.
 The frequency-domain channel at the $k$th subcarrier, $k=1,\cdots,K$, can be expressed as  \cite{alkhateeb2016frequency}
 \begin{equation}\label{eq:subcarrier channel}
     \Hb_k=\sum_{d=0}^{D-1} \Hb_{f_k}[d] e^{-j\frac{2\pi k}{K}d},
 \end{equation}
 where $f_k$ denotes the central frequency of the $k$th subcarrier with bandwidth $B$, which can be described as \cite{li2020dynamic}
  \begin{equation}\label{eq:carrier frequency}
    f_{k}=f_c+\left(k-\frac{K+1}{2} \right)\frac{B}{K}, \quad \forall k.
  \end{equation}

\subsection{Problem Formulation}
Assuming the availability of full channel state information, we aim at designing the combiner of SW-HBF that maximizes the system SE of the considered wideband MIMO-OFDM system. For the PS-HBF (see Fig. \ref{fig: PS HBF}), the set of feasible analog combining vectors (i.e., the columns of $\Wb_{RF}$) is given by $\Uc_1=\left\{\ub \in \Cs^{N_r \times 1} \left| |u_j|=1 , j=1,\cdots, N_r \right. \right\}$, where $u_j$ denotes the $j$th element of vector $\ub $, $|x|$ denotes the modulus of a complex number $x$. Whereas the feasible set of SW-based analog combiner in the SW-HBF scheme (see Fig. \ref{fig: SW HBF}) is given by  $\Uc_2=\left\{\ub \in \Bc^{N_r \times 1}  \right\}$, where $\Bc=\{0,1\}$. Assuming that the transmit symbol at each subcarrier follows a Gaussian distribution, the problem of designing the combiner in the PS-HBF and SW-HBF schemes can be formulated as
\begin{subequations}\label{eq:problem formulation}
\begin{align}
  \underset{\Wb_{RF},\atop \left\{ \Fb_k, \Wb_{BB}[k]\right\}_{k=1}^{K}}{\max}  &  \frac{1}{K}\sum_{k=1}^{K}\log_2\left|\Ib_{N_s}+\frac{1}{\sigma^2_n}\Wb_k^{\dagger}\Hb_k\Fb_k\Fb^H_k\Hb^H_k\Wb_k\right|  \\
  \rm{s.t.} \qquad & \quad\sum_{k=1}^{K}\|\Fb_k\|^2_F \leq P_b,  \\
  &  \quad \Wb_{RF}(:,j) \in \Uc, \, j=1,\cdots,N_{RF}, \label{st:pb binary constraint} \\
  & \quad \rank( \Wb_{RF})\geq N_s \label{st:pb rank constraint},
\end{align}
\end{subequations}
where $\dagger$ denotes the Moore-Penrose pseudo inversion, $\Wb_{RF}(:,j)$ denotes the $j$th column of $\Wb_{RF}$ and $P_b$ is the transmit power budget. Note that $\Uc=\Ac_1$ for PS-HBF; in contrast, for SW-HBF, $\Uc=\Ac_2$. 

\section{SW-HBF Design}
The PS-HBF design problem can be solved by the methods proposed in \cite{ma2021closed,sohrabi2017hybrid}. In contrast, SW-HBF design is more challenging, and its solution is unavailable in the literature. In \eqref{eq:problem formulation}, the objective function is non-convex due to the binary variable constraints \eqref{st:pb binary constraint} and the rank constraint \eqref{st:pb rank constraint} of the analog combiner. The optimal solution can be found by the exhaustive search, which, however, is computationally prohibitive. To overcome this, we develop a computationally efficient solution to the SE maximization problem by decoupling the design of $\left\{ \Fb_k \right\}_{k=1}^{K}$ and  $\left\{ \Wb_k \right\}_{k=1}^{K}$. Specifically, for the design of transmit beamformers $\left\{ \Fb_k \right\}_{k=1}^{K}$, we assume that the optimal receiver is used. Then, the receive beamformers $\left\{ \Wb_k \right\}_{k=1}^{K}$ are obtained given the transmit beamformers $\left\{ \Fb_k \right\}_{k=1}^{K}$. The solutions to these subproblems are presented in the following subsections. 

\subsection{Transmit Beamformer Design}
Given the receive beamforming matrices $\left\{ \Wb_k \right\}_{k=1}^{K}$, the transmit beamforming design problem is expressed as
\begin{align}\label{eq:problem of OFDM transmitter}
  \max\limits_{\{\Fb_k\}_{k=1}^{K}} & \quad \frac{1}{K}\sum_{k=1}^{K}\log_2\left|\Ib_{N_r}+\frac{1}{\sigma_n^2}\Hb_k\Fb_k\Fb^H_k\Hb^H_k\right|  \\
  \text{s.t.} & \quad\sum_{k=1}^{K}\|\Fb_k\|^2_F \leq P_b \notag.
\end{align}
 Let $\Hb_k=\Ub_{k}\SigmaB_{k} \Vb_{k}^H$ be the singular value decomposition (SVD) of $\Hb_k$, where $\Ub_{k}\in \Cs^{N_r \times N_s}, \SigmaB_{k}\in \Cs^{N_s \times N_s}, \Vb_{k}\in \Cs^{N_t \times N_s}$. The optimal solution for $\Fb_k$ can be given as \cite{zhang2005variable}
\begin{equation}\label{eq:solution of OFDM transmitter}
  \Fb_k=\Vb_{k}\Gammab_k^{\frac{1}{2}}\Bb_k,
\end{equation}
where $\Bb_k$ is any $N_s\times N_s$ unitary matrix. $\Gammab_k=\diag(p_{k,1},\cdots,p_{k,N_s})$ (satisfying $\sum_{k=1}^{K} Tr(\Gammab_k)=P_b$) is the diagonal matrix obtained by the water-filling approach, i.e.,
  \begin{equation}\label{eq:water-filling solution for single carrier}
    p_{k,i}=\left[\mu-\frac{\sigma_n^2}{\lambda_{\Hb_k,i}}\right]^+,
  \end{equation}
  where $\lambda_{\Hb_k,i}$ denotes the $i$th largest eigenvalue of $\Hb_k$, $\mu$ is the water level, and in \eqref{eq:water-filling solution for single carrier},  $[x]^+ \triangleq \max(x,0)$, $\forall x \in \Rs$. 

\subsection{Receive Beamformer Design}
For a fixed analog combiner $\Wb_{RF}$, the optimal digital combiner of each subcarrier is the MMSE solution \cite{ma2021closed},
 \begin{equation}\label{eq:optimal baseband combiner}
   \Wb_{BB}[k]=\left(\Jb_k\Jb^H_k+\sigma_n^2\Wb_{RF}^H\Wb_{RF}\right)^{-1}\Jb_k,
 \end{equation}
where $\Jb_k\triangleq \Wb_{RF}^H \Hb_k\Fb_k$. Since the optimal MMSE digital combiner can achieve maximum SE , the problem of designing analog combiner is expressed as \cite{ shi2018spectral}
\begin{subequations}\label{eq:pb for combiner design}
\begin{align}
  \underset{\Wb_{RF}}{\max}  & \quad \frac{1}{K}\sum_{k=1}^{K}\log_2\left|\Ib_{N_{RF}}+\frac{1}{\sigma^2_n}\Wb_{RF}^{\dagger}\Tilde{\Fb}_k\Wb_{RF}\right|  \label{goal:objective functio for designing combiner}\\
  \rm{s.t.} &  \quad \Wb_{RF}(:,j) \in \Uc_2, \, j=1,\cdots,N_{RF}, \label{st:pb binary constraint for combiner} \\
  & \quad \rank( \Wb_{RF})\geq N_s \label{st:pb rank constraint for combiner},
\end{align}
\end{subequations}
where $\Tilde{\Fb}_k\triangleq \Hb_k\Fb_k\Fb^H_k\Hb^H_k$. Problem \eqref{eq:pb for combiner design} is still challenging due to the binary variable constraints \eqref{st:pb binary constraint for combiner} and rank constraint \eqref{st:pb rank constraint for combiner}. The optima can be found by exhaustive search, which is prohibitively complex. To solve the problem efficiently, we develop a low complexity algorithm based on tabu search (TS) method \cite{nguyen2019qr}. 

TS algorithm begins to search for the neighbors of an initial point $\wb_0$ and records the best neighbor with the largest objective value $f(\wb_b)$ as the best candidate $\wb_b$. It then iteratively searches for the neighbors of the best candidate $\wb_b$ and updates $\wb_b$ until the stopping criteria are met. During the process, a tabu list $\Lc$ of length $L$ is used to record the visited points to avoid cycling. The output of the TS algorithm is the best solution $\wb_b^*$ found in the iterations, which achieves the largest value of the objective function. 

For problem \eqref{eq:pb for combiner design}, the objective function can be defined as
\begin{equation}\label{eq:fitness function}
    f(\Wb_{RF})=\frac{1}{K}\sum_{k=1}^{K}\log_2\left|\Ib_{N_{RF}}+\frac{1}{\sigma^2_n}\Wb_{RF}^{\dagger}\Tilde{\Fb}_k\Wb_{RF}\right|.
\end{equation}
For notation convenience, we will use $\Wb$ in the following part to represent $\Wb_{RF}$. Let $\eb=[e_1,\cdots,e_{N_r N_{RF}}]^T  \in \Bc^{N_r N_{RF}}$, and $\eb_i$ be the vector in which $i$th element is $1$ and other elements are $0$, i.e., $e_i=1,e_{j \neq i}=0, \forall j$. Let $\Wb_b$ be the found best candidate. Furthermore, let  $\wb_b\triangleq \vectorize(\Wb_b)$ be the vectorization of $\Wb_b$. The neighbor set of $\wb_b$ is given by
\begin{align}
    \Nc(\wb_b)=&\{\wb \in \Bc^{N_r N_{RF}} \left| |\wb-\wb_b|=\eb_i,\rank(\Wb) \geq N_s, \right.\nonumber\\
    & i=1,\cdots, N_r N_{RF}  \},
\end{align}
where $ \Wb \triangleq \vectorize^{-1}(\wb)$, $|\xb|$ ($\xb\in \Bc^{N_r N_{RF}\times 1}$) denotes the vector which has element-wise absolute value of $\xb$. The TS algorithm solving problem \eqref{eq:pb for combiner design} is summarized in Algorithm \ref{alg:TS algorithm} where $\Wb_0$ is the initial candidate, and $N_{iter}$ is the maximum iteration.
\begin{algorithm}[hbpt]
\caption{TS algorithm for solving problem \eqref{eq:pb for combiner design}}\label{alg:TS algorithm}
\LinesNumbered 
\KwIn{$L,\Wb_0,N_{iter},i=0$}
\KwOut{$\Wb_b^*=\vectorize^{-1}(\wb_b^*)$}

$\wb_0=\vectorize(\Wb_0),\wb_b \gets \wb_0, \Wb_b \gets \Wb_0, \wb_b^* \gets \wb_0 $\;
$\Lc =\varnothing,\Lc \gets \Lc \cup \wb_0$ \;
\While{$i \leq N_{iter} \; \&$  {\rm not converge}}{

\For{$\wb \in \Nc(\wb_b) $ }
{
$\Wb=\vectorize^{-1}(\wb)$\;
    \If{$\wb \notin \Lc \, \& \, f(\Wb) > f(\Wb_b) $ }
    {$\wb_b \gets \wb$\;}
}
$\Wb_b=\vectorize^{-1}(\wb_b),\Wb_b^*=\vectorize^{-1}(\wb_b^*)$\;
    \If{$ f(\Wb_b) > f(\Wb_b^*) $ }
    {$\wb_b^* \gets \wb_b$\;}
    $\Lc \gets \Lc \cup \wb_b$\;
    
    \If{$ |\Lc|> L $ }
    {Remove the earliest $\wb_b$ in $\Lc$\;}
    $i \gets i+1$\;
}
\end{algorithm}
\begin{remark}
The convergence of the TS algorithm is guaranteed by iteratively moving to the best neighbor with an equal or larger objective value from an initial point. Therefore, it has the potential to find the near-optimal solution, which can be corroborated by the results shown in Section \ref{sc:results}.
\end{remark}

In Algorithm \ref{alg:TS algorithm}, line 1 initializes the best candidate as the chosen point $\Wb_0$, which is then added to the tabu list (line 2). Based on the initial candidate $\Wb_0$, the procedure iteratively searches for the best neighbor and updates the best candidate until convergence (lines 3-19). In each iteration, the TS algorithm first collects the neighbor set $\Nc(\wb_b)$ and treats each neighbor $\wb \in \Nc(\wb_b)$ as the potential candidate. By comparing the objective value of all potential candidates to that of the previous best candidate, the new best candidate, which is not in the tabu list, is found (see lines 4-9). Afterward, the procedure updates the best solution $\wb_b^*$ via comparing it with the best candidate $\wb_b$ (lines 10-13). Finally, the best candidate $\wb_b$ is added to the tabu list to avoid repeated cycling of future search (lines 14-17). 

Since the TS algorithm adopts the local search procedure, the initial search point can greatly impact its performance and computational complexity. Thus, we further develop a heuristic algorithm to improve the quality of the initial point of the TS algorithm. By removing the rank constraint \eqref{st:pb rank constraint for combiner} and relaxing the binary variable constraints  \eqref{st:pb binary constraint for combiner}, the problem is recast as
\begin{align}\label{eq:pb initial point}
  \underset{\Wb}{\max}  & \quad f(\Wb)= \frac{1}{K}\sum_{k=1}^{K}\log_2\left|\Ib+\frac{1}{\sigma^2_n}\Wb^{\dagger}\Tilde{\Fb}_k\Wb\right|  \\
  \rm{s.t.} &  \quad \Wb \in \Wc  \nonumber,
\end{align}
where $\Wc=\{\Wb| \Wb(i,j)\in [0,1], \forall i,j \} $, $\Wb(i,j)$ denotes the element of $\Wb$ at $i$th row and $j$th column. Given an arbitrary initial point, problem \eqref{eq:pb initial point} can be efficiently solved by the projected gradient ascending (PGA) algorithm, which is summarized in Algorithm 2. $[\cdot]_{\Wc}$ denotes the projection into $\Wc$.
\begin{table}[htbp]
  \vspace{-0.2cm}
  \centering
  \begin{tabular}{cl} 
    \hline
    &{{\bf Algorithm 2}: PGA algorithm for problem \eqref{eq:pb initial point}}\\
    \hline
    &1. Initialize: $i=1,\Wb^i\in \Wc,c=1$\\
    &2. \quad Repeat.\\
    &3. \qquad  $\alpha=\frac{c}{\sqrt{i+1}}$. \\
    &4. \qquad $\mathbf{W}^{i+1} \leftarrow [\mathbf{W}^{i}+\alpha\nabla_{\Wb^i}f(\Wb^i)]_{\Wc}$.\\
    &5. \qquad $i=i+1$.\\
    &6. \quad Until convergence.\\
    &7.  Output: $\Wb_{pga}$.\\
    \hline
  \end{tabular}
  \vspace{-0.3cm}
\end{table}

Let $\Wb_{pga}$ denote the output of the PGA algorithm. The initial search point of the TS algorithm can be obtained by rounding $\Wb_{pga}$ to the nearest solution in the feasible space of problem \eqref{eq:pb for combiner design}, i.e.,
\begin{equation}
    \Wb_0=r(\Wb_{pga}),
\end{equation}
where $r(\cdot)$ is the rounding function. By integrating Algorithm 2 into the Algorithm \ref{alg:TS algorithm}, we get an improved TS algorithm, which is termed as {\it PGA-aided TS algorithm}.

\section{Power consumption model}
Based on the architectures illustrated in Fig. \ref{fig: PS HBF} and \ref{fig: SW HBF}, the total power consumption of the fully digital beamforming (DBF), PS-HBF, and SW-HBF schemes are given as
\begin{align}\label{eq:architecture power model}
    P^{\rm DBF}_{\rm total}&=N_r(P_{LNA}+P_{RF}+2P_{ADC}),\\
    P^{\rm PS-HBF}_{\rm total}& =N_r(P_{LNA}+P_{SP}+N_{RF}P_{PS}) \nonumber\\
  &\qquad\qquad +N_{RF}(P_{RF}+P_C+2P_{ADC}), \\
  P^{\rm SW-HBF}_{\rm total} &=N_r(P_{LNA}+P_{SP}+N_{RF}P_{SW}) \nonumber\\
 &\qquad\qquad +N_{RF}(P_{RF}+P_C+2P_{ADC}),
\end{align}
respectively, where $P_{RF}$ represents the power consumption of an RF chain, which can be given as
\begin{equation}\label{eq:RF power model}
   P_{RF}=P_{M}+P_{LO}+P_{LPF}+P_{BBamp} , 
\end{equation}
where $P_{M}$, $P_{LO}$, $P_{LPF}$, and $P_{BBamp}$ are the power consumption of the mixer, the local oscillator, the low pass filter, and the baseband amplifier, respectively. As a result, the system EE is defined as
\begin{equation}
    EE=\frac{SE}{P_{\rm total}},
\end{equation} 
where $EE$ and $SE$ represent the energy efficiency and spectral efficiency, respectively.
\section{Simulation Results}\label{sc:results}
In this section, we present the numerical results to evaluate the performance and computational complexity of the proposed SW-HBF schemes in the considered system. In the simulations, we use the channel model given in Section \ref{sc:channel model} with $L=10, d_s=\frac{\lambda_c}{2},f_c=60{\rm GHz},B=1{\rm GHz}, K=64$. The AoA/AoDs are uniformly distributed over $[0,2\pi)$, and the pulse shaping filter is modeled as \cite{alkhateeb2016frequency}
\begin{align*}
    p(t)=
    \begin{cases}
        \frac{\pi}{4} \operatorname{sinc}\left(\frac{1}{2 \beta}\right), &\text{if~} t=\pm \frac{T_{\mathrm{s}}}{2 \beta} \\
        \operatorname{sinc}\left(\frac{t}{T_{\mathrm{s}}}\right) \frac{\cos \left(\frac{\pi \beta t}{T_{\mathrm{s}}}\right)}{1-\left(\frac{2 \beta t}{T_{\mathrm{s}}}\right)^{2}}, &\text {otherwise},
    \end{cases}
    \nthis \label{eq:pulse shaping}
\end{align*}
with $T_s$ the sampling period and the roll-off factor $\beta=1$. The path delay is uniformly distributed in $[0,(D-1)T_s]$ where $D$ is the cyclic prefix length, given by $D=K/4$ according to 802.11ad. The SNR is defined as SNR$\triangleq \frac{P_b}{K\sigma_n^2}$. The assumptions on the component power consumptions are given in the Table \ref{Tb:power each device} \cite{mendez2016hybrid,abbas2017millimeter}. All reported results are averaged over $10^3$ channel realizations.
\begin{table}[htbp]
    \centering
    \caption{Power consumption of each device}\label{Tb:power each device}
        \begin{tabular}{rlr}
            \hline
            Device &  Notation   & Value  \\
            \hline
            Low Noise Amplifier (LNA) & $P_{LNA}$ &39mW\\
            Splitter         & $P_{SP}$ &19.5mW\\
            Combiner        & $P_C$     &19.5mW\\
            Phase shifter    &$P_{PS}$  & 30mW\\
            Switch          & $P_{SW}$ & 5mW\\
            Mixer            & $P_M$     & 19mW\\
            Local oscillator  & $P_{LO}$   & 5mW\\
            Low pass filter   & $P_{LPF}$  & 14mW\\
            Base-band amplifier & $P_{BBamp}$ & 5mW\\
            ADC              & $P_{ADC}$   & 240mW\\
            \hline
        \end{tabular}
        \vspace{-0.3cm}
\end{table}
\subsection{Performance Evaluation}
\begin{figure}[htbp]
   \vspace{-0.3cm}
        \centering	
         \includegraphics[width=0.4\textwidth]{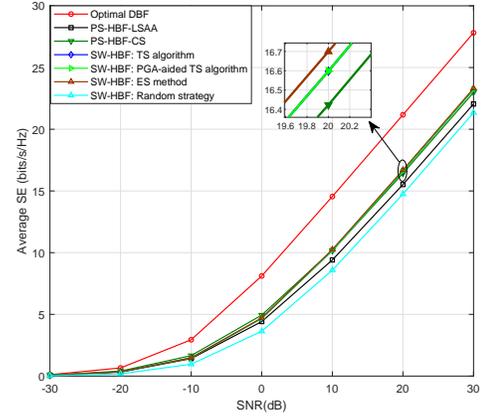}
         \captionsetup{font={small}}
        \caption{SE of considered HBF algorithms vs SNR with $N_t=16,N_r=8,N_s=N_{RF}=2,K=64$}
        	\label{fig:SE of different HBF algorithms vs SNR}
      \vspace{-0.3cm}
  \end{figure}

\begin{figure}[htbp]
   \vspace{-0.3cm}
        \centering	
         \includegraphics[width=0.4\textwidth]{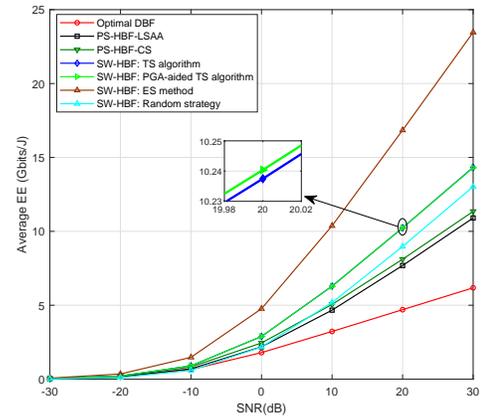}
         \captionsetup{font={small}}
        \caption{EE of the considered HBF algorithms vs SNR with $N_t=16,N_r=8,N_s=N_{RF}=2,K=64$}
        	\label{fig:EE of different HBF algorithms vs SNR}
      \vspace{-0.3cm}
  \end{figure}

Figs.\ \ref{fig:SE of different HBF algorithms vs SNR} and \ref{fig:EE of different HBF algorithms vs SNR} show the average SE and EE of different algorithms versus the SNR. For comparison, we include the PS-HBF in large-scale antenna arrays (PS-HBF-LSAA) algorithm in \cite{sohrabi2017hybrid}, and the PS-HBF with closed-form solutions (PS-HBF-CS) in \cite{ma2021closed}. For SW-HBF, the performance of optimal solution obtained by exhaustive search (ES) method and randomly generated solution by the random strategy are presented. Moreover, the performance of optimal digital beamforming (DBF) via the water-filling algorithm is also exhibited. It can be observed from Figs.\ \ref{fig:SE of different HBF algorithms vs SNR} and \ref{fig:EE of different HBF algorithms vs SNR} that the optimal SW-HBF can achieve the best performance in terms of SE and EE. Furthermore, the TS and PGA-aided TS algorithms can obtain near-optimal solutions and perform better than the PS-HBF-LSAA and PS-HBF-CS algorithms. The performance of the random solution is considerably worse than the TS algorithm, which demonstrates the effectiveness of the proposed TS algorithms. Finally, we can observe that the performance of the PGA-aided TS algorithm is slightly better than that of the vanilla TS algorithm. Based on the results shown in Figs.\ \ref{fig:SE of different HBF algorithms vs SNR} and \ref{fig:EE of different HBF algorithms vs SNR}, we can conclude that the SW-HBF scheme is able to provide better SE and EE performance than that of PS-HBF schemes in wideband multi-carrier systems.

\begin{figure}[htbp]
  \vspace{-0.5cm}
        \centering
        \subfigure[SE vs bandwidth.]
        {\label{fig:SE versus bandwidth} \includegraphics[width=0.23\textwidth]{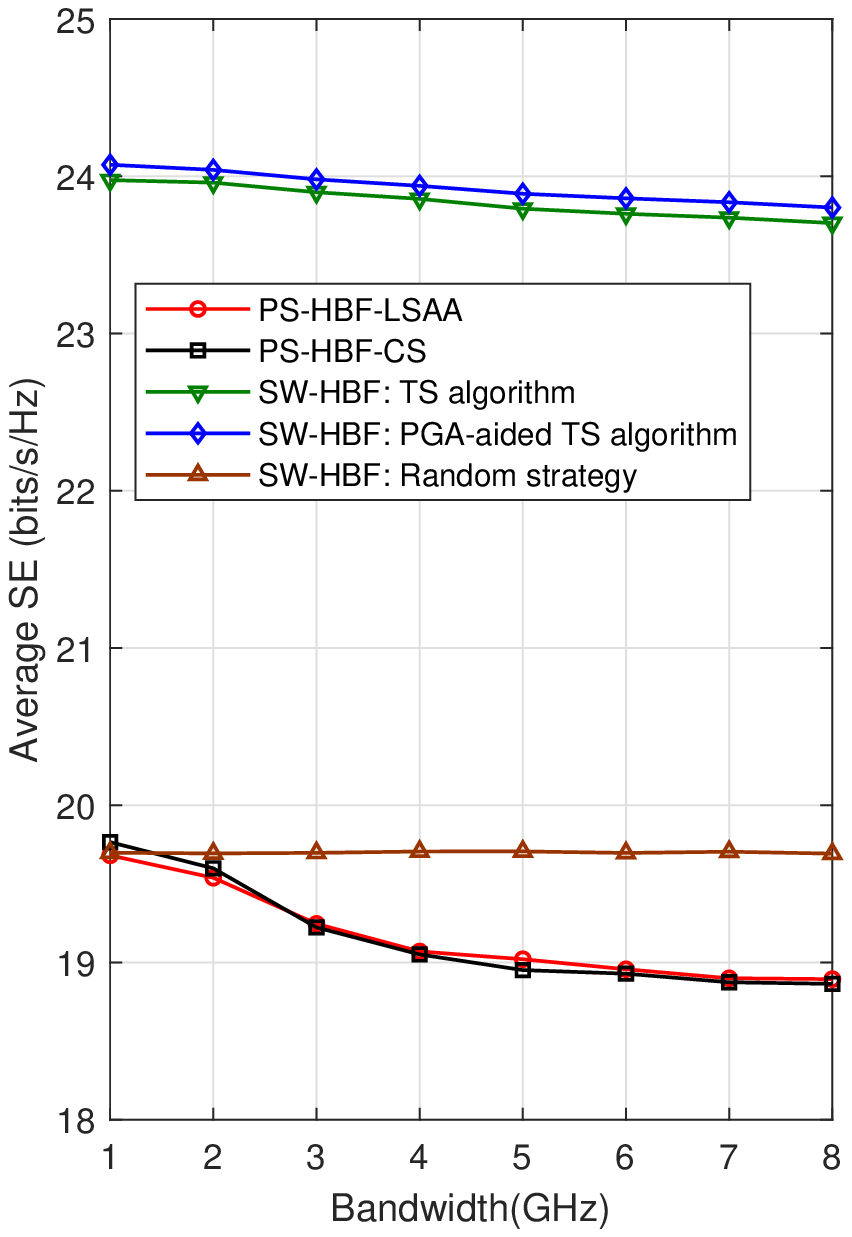}}
        \subfigure[SE vs the number of subcarriers.]
        {\label{fig:SE versus the number of subcarriers}\includegraphics[width=0.23\textwidth]{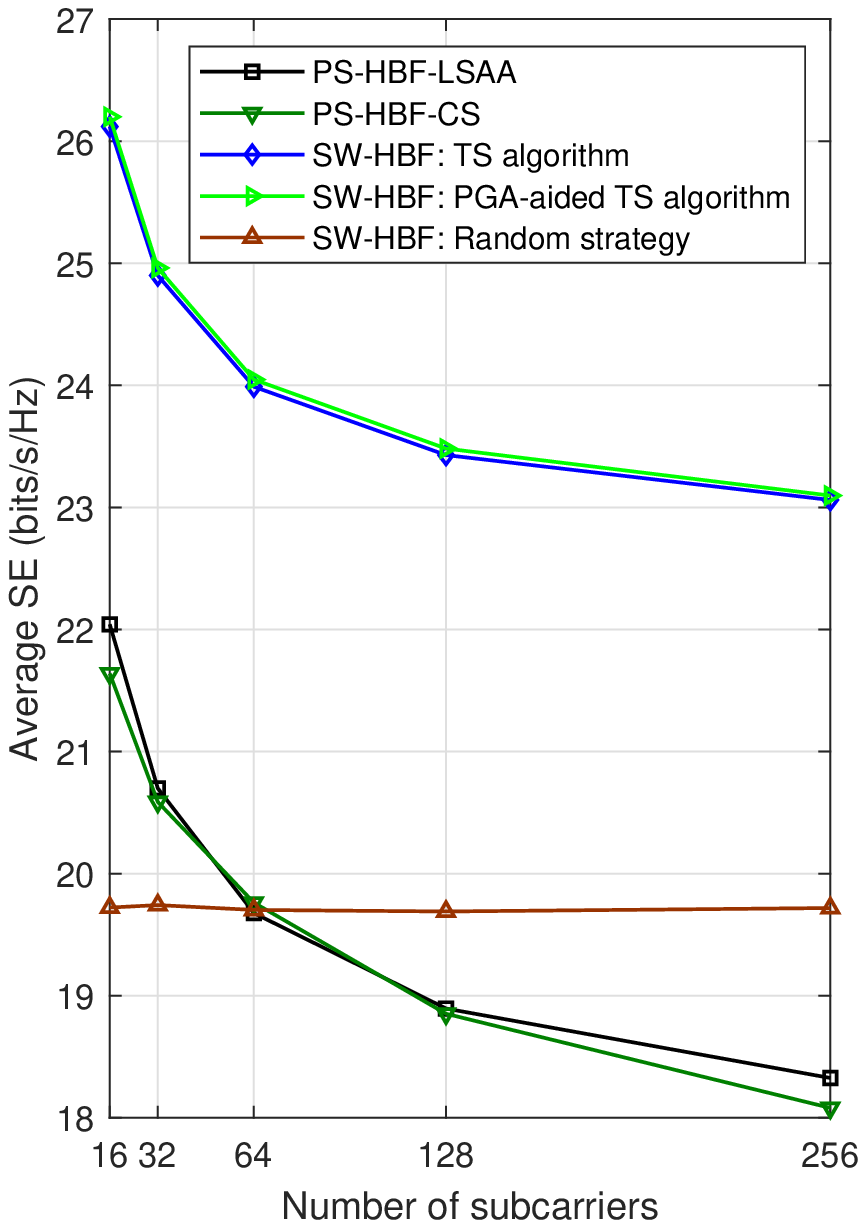}}
        \captionsetup{font={small}}
        \caption{SE versus system bandwidth and number of subcarriers with $N_t=64,N_r=64,N_s=N_{RF}=4,f_c=60{\rm GHz}$.}
        \label{fig:SE versus system bandwidth and number of subcarriers}
        	\vspace{-0.3cm}
  \end{figure}

Figs.\ \ref{fig:SE versus bandwidth} and \ref{fig:SE versus the number of subcarriers} show the system SE versus the system bandwidth and the number of subcarriers, respectively. We can observe from Fig.\ \ref{fig:SE versus bandwidth} that as the bandwidth increases, the system suffers from more severe beam squint effect, rendering a significant loss in the SE. With the proposed TS algorithms, the SW-HBF is less affected by the beam squint effect and can achieve higher SE than PS-HBF. Moreover, it can be observed from Fig.\ \ref{fig:SE versus the number of subcarriers} that with more subcarriers, the use of the common analog beamformer induces larger loss of SE. However, when there are more subcarriers, the channel correlation of subcarriers gets larger, making the system SE less affected by the analog beamformer. This explains the slight decrease of the SE with the increasing number of subcarriers, as is shown in Fig.\ \ref{fig:SE versus the number of subcarriers}. In summary, the TS-based SW-HBF schemes are less affected by the beam squint effect and the number of subcarriers, and they can achieve higher SE than the PS-HBF-LSAA and PS-HBF-CS schemes. Moreover, the performance of the PGA-aided TS algorithm is slightly better than that of the TS algorithm without a PGA solution.
  \subsection{Complexity Analysis}
\begin{figure}[htbp]
  \vspace{-0.3cm}
        \centering
        \subfigure[Comparison of computational complexity.]
        {\label{fig:complexity1} \includegraphics[width=0.23\textwidth]{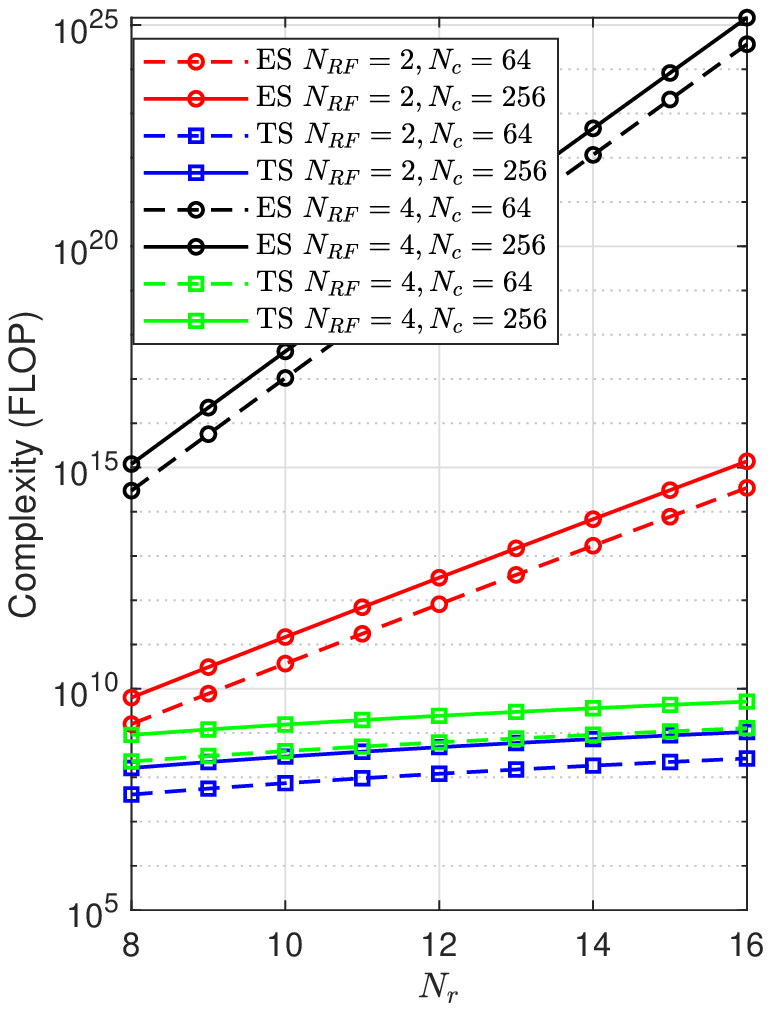}}
        \subfigure[Computational complexity of TS algorithm.]
        {\label{fig:complexity2}\includegraphics[width=0.23\textwidth]{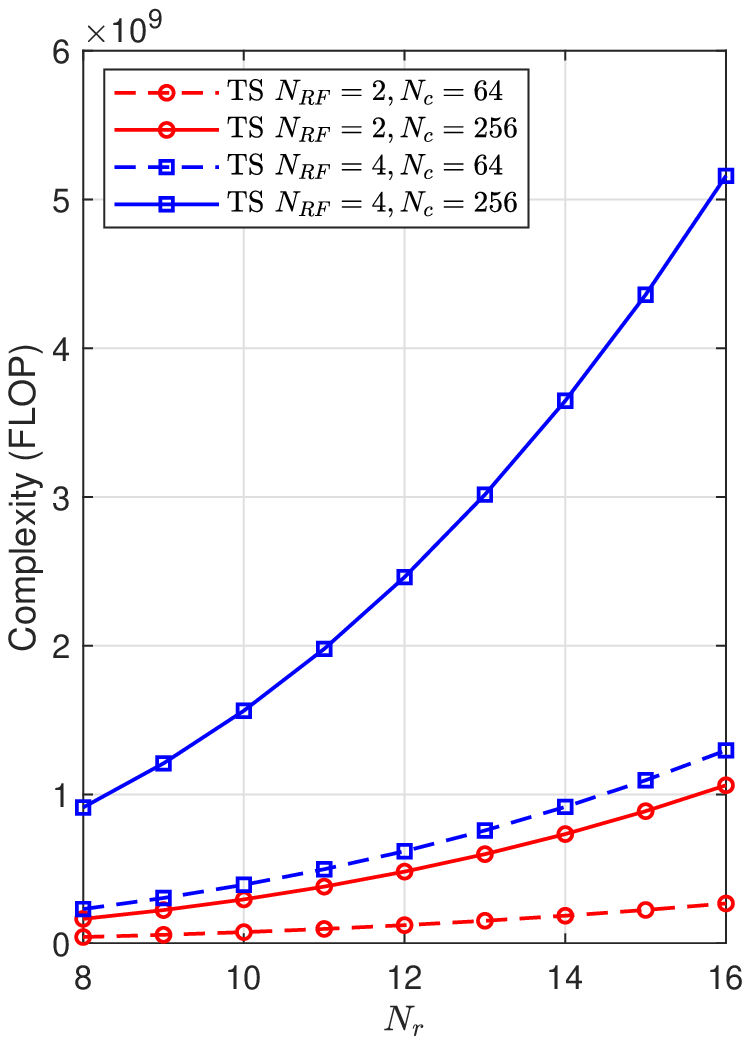}}
        \captionsetup{font={small}}
        \caption{Complexity analysis.}
        \label{fig:Complexity analysis.}
        	\vspace{-0.3cm}
  \end{figure}

  The complexity of the ES method and proposed TS algorithms come from the computation of the objective function \eqref{goal:objective functio for designing combiner} and the rank of the potential solution $\Wb$. Since the TS procedure dominates the complexity of the PGA-aided TS algorithm, the complexities of the two considered TS algorithms are approximately the same. The complexity of the ES method and the TS algorithm are $\Oc(K N_r^2 N_{RF}2^{N_r N_{RF}})$ and $\Oc(N_{iter} K N_r^3 N_{RF}^2)$, respectively. As shown in Figs.\ \ref{fig:complexity1} and \ref{fig:complexity2}, the proposed TS algorithms have much lower computational complexity compared with the optimal ES method.
  
  \section{Conclusion}
  In this paper, we study the performance of SW-HBF in a MIMO-OFDM system with the frequency-selective wideband channel. The near-optimal solution to the analog combiner that maximizes the system SE is obtained via the proposed two TS algorithms. Furthermore, we present the power consumption model of the SW-HBF and PS-HBF architectures. Numerical simulations compare the SE and EE achieved by the SW-HBF to those of the PS-HBF schemes. They demonstrate that the former is able to obtain better SE and EE and less affected by the beam squint effect than the latter. The results show that employing SW-HBF can reap more benefits than using PS-HBF in a wideband multi-carrier system. 
%
\IEEEpeerreviewmaketitle

\bibliographystyle{IEEEtran}
\bibliography{conf_short,jour_short,SW_HBF}

\end{document}